\documentclass[twocolumn,showpacs,preprintnumbers,amsmath,amssymb]{revtex4}
\usepackage{graphicx}
\usepackage{dcolumn}
\usepackage{bm}
\begin{document}

\title{Pressure Induced \textit{Complexity} in a Lithium Monolayer}

\author{A. Rodriguez-Prieto}

\author{A. Bergara}
\affiliation{Materia Kondentsatuaren Fisika Saila, Zientzia eta Teknologia Fakultatea, Euskal Herriko 
Unibertsitatea, 644 Postakutxatila, 48080 Bilbo, Basque Country, Spain}

\date{\today}

\begin{abstract}
Light alkali metals have usually been considered as \textit{simple} metals due to their monovalency and high 
conductivity. In these metals ionic pseudopotentials are weak and the nearly free electron model (NFE) 
becomes quite accurate at normal conditions. However, very recent experiments have shown that at high pressures
their electronic properties deviate radically from the NFE model and even become unexpected good superconductors. 
In this work we present {\it ab initio} calculations to analyze the deviation from simplicity in a lithium
monolayer (ML) when pressure is applied. We have seen that as a result of the increasing non-local character
of the atomic pseudopotential with increasing pressure, the surprising half filling Hubbard-type nesting observed 
in the Fermi {\it line} can explain the interesting \textit{complex} behavior in lithium ML, induced by its correlated structural, electronic and even magnetic properties.

\end{abstract}

\maketitle

\section{Introduction}

Light alkalies are usually considered as \textit{simple} metals, because as a consequence of 
the weak interaction between valence 
electrons and ionic cores, under normal conditions they crystallize in simple high symmetric structures. Therefore, 
the nearly free electron model (NFE) has been considered a good approximation to describe their electronic 
properties \cite{wigner}. The only exception is hydrogen which, even when solidifies, forms very stable diatomic molecules and presents an insulating character. However, recent both theoretical and experimental results have changed our perception of the alkalies and have shown that the NFE model breaks when high pressures are applied. Neaton {\it et al.} reported {\it ab initio} calculations for bulk lithium \cite{NA} and predicted that pressure could induce structural transitions to less symmetric, lower coordinated structures, associated with electronic localizations. These theoretical predictions have been confirmed by experiments \cite{hanfland}, finding that around 40 GPa a phase transition to a complex structure ($cI$16) with 16 atoms per unit cell occurs.
It is important to notice that pressure induced  transitions from simple to more complex structures are 
not singular to lithium but have also been predicted in heavier alkalies \cite{NeatonNa, ChristensenNa}.
On the other hand, other very recent experiments \cite{shimizu,str,deemyad} have also shown that at 
this pressure range lithium presents 
a superconducting transition with T$_c\sim 20$K, becoming the highest transition temperature between
simple elements \cite{ashcroft_nature}. It is noteworthy that experiments looking for superconductivity
in lithium under ambient pressure have failed \cite{liambient}. This even rises the interest to characterice
the physical properties of compressed lithium close to the observed electronic and structural transitions.\\

In this article we present an {\it ab initio} theoretical analysis of the pressure-induced structural, electronic 
and even magnetic properties of a lithium ML. The development of new techniques for the atomic manipulation
\cite{14,15} allows the growth of atomic monolayers on inert substratum, semiconductors or noble gases. These new
possibilities rise the interest to analyze physical properties of low dimensional systems under different conditions.
On the other hand, the simplicity of the atomic configuration associated to the ML facilitates the performing of a
more detailed analysis, and extending our conclusions to the bulk will also provide
an interesting perspective to understand the physical origin of the experimentally observed features of compressed lithium. In Section \ref{sec:T} we describe the theoretical and computational background 
of this work. Results and Conclusions will be presented in Sections \ref{sec:R} and \ref{sec:C}, 
respectively.\\

\section{Theoretical and Computational Background}
\label{sec:T}
In order to describe structural and electronic properties of a lithium ML over a wide range of densities 
we have used a plane-wave implementation of the Density Functional Theory (DFT)
\cite{dft1,dft2} 
within the Local Density Approximation (LDA) \cite{lda1,lda2,lda3} for non magnetic calculations and 
Local Spin Density Approximation (LSDA) \cite{lsda} to analyze the magnetic properties of the ML. All 
calculations have been performed with the Vienna Ab initio Simulation Package (VASP)\cite{vasp1,vasp2}. 
Effects of compression were simulated by reducing the lattice parameter. 
As the atomic spacing is reduced a significant core overlap is observed, so that
we have required methods which take into account core electrons. For fully treating them we 
make use of the Projector Augmented Wave (PAW) method \cite{paw}.
A Monkhorst-Pack \cite{MP} mesh of $20\times 20\times 1$ has been chosen for the sampling of the Brillouin zone. 
The description of the ML is implemented by a supercell which, in order to minimice interactions
between MLs, contains ten layers of vacuum between MLs. The ions are relaxed 
to their equilibrium positions by calculating the forces acting on them using 
the Hellmann-Feynmann theorem. In all the calculations we have used  an energy cutoff of 
$E_{\rm cut}\simeq 815 \mathrm{eV}$.\\

\section{Results}
\label{sec:R}
\subsection{Structural and Electronic Analysis.}
\label{subsec:monoatomic}

\begin{figure}[h!]
\includegraphics[width=\linewidth,clip=true]{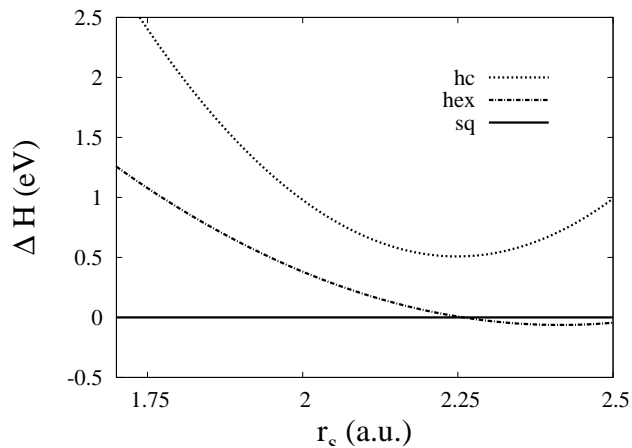}
\caption{ Enthalpy $H=E+\sigma A$, $\sigma$ being the surface tension: $\sigma=-dE/dA$, as a function of the density parameter, $r_s$, for three simple structures: hexagonal ($hex$), square ($sq$) and
honeycomb ($hc$). All enthalpies are referenced to the square structure, $\Delta H=H-H_{sq}$.}  
\label{fig:Fig1}
\end{figure}

Following the analysis in Ref. \cite{aitor}, as a first step we will just consider three simple structures 
of a lithium ML: hexagonal ($hex$), square ($sq$) and honeycomb ($hc$). 
Fig. \ref{fig:Fig1} shows the calculated enthalpies for these structures. 
As it is a common property of the alkalies, lithium ML at equilibrium ($r_s=3.02$ a.u.) \cite{rs} also favors the compact hexagonal structure ($hex$). However, as we can observe in Fig.\ref{fig:Fig1}, at $r_{s}< 2.25$ a.u. the more open square structure ($sq$) becomes favored over the hexagonal one.

\begin{figure}[h!]
\includegraphics[width=\linewidth,clip=true]{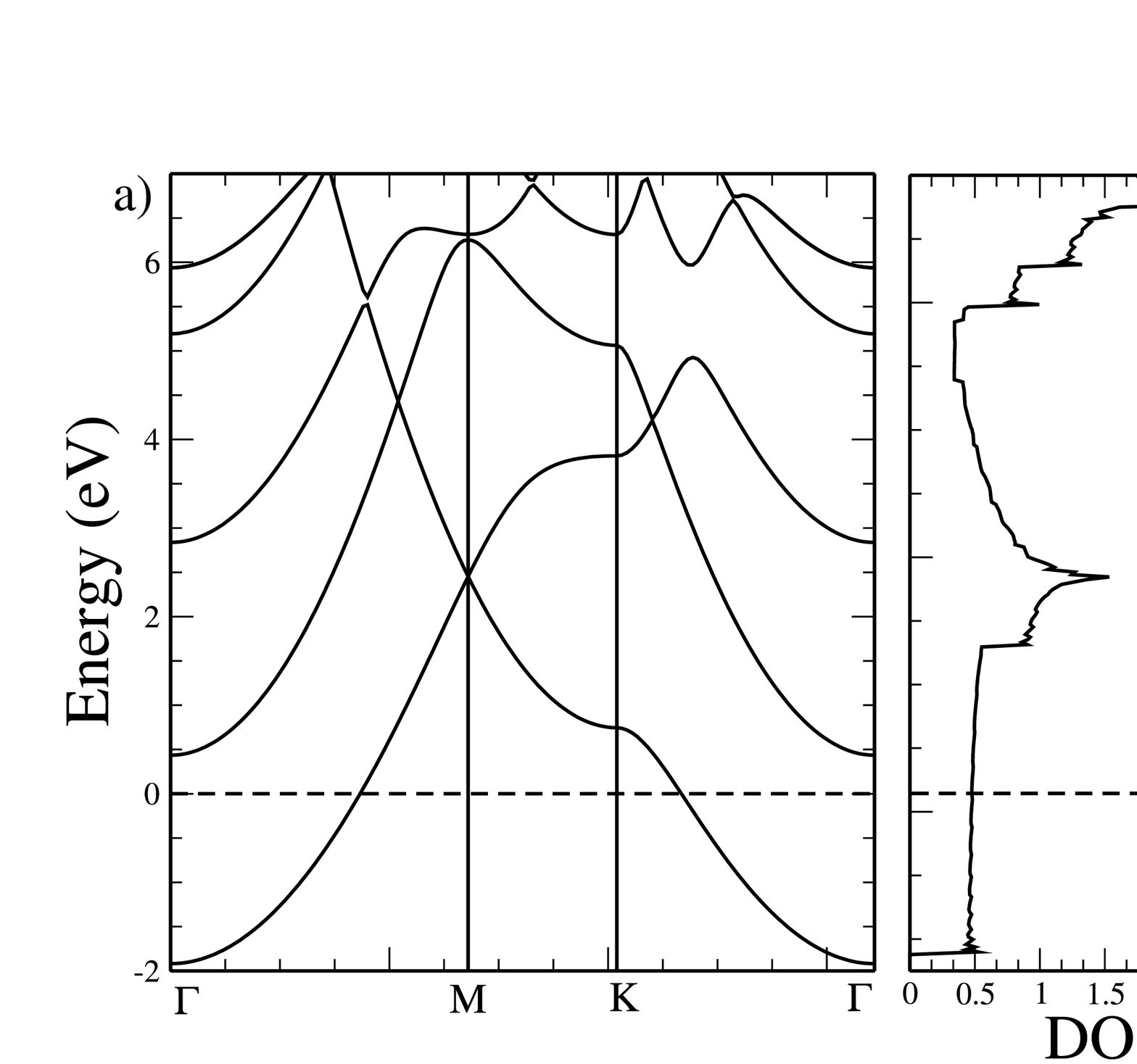}\\
\includegraphics[width=\linewidth,clip=true]{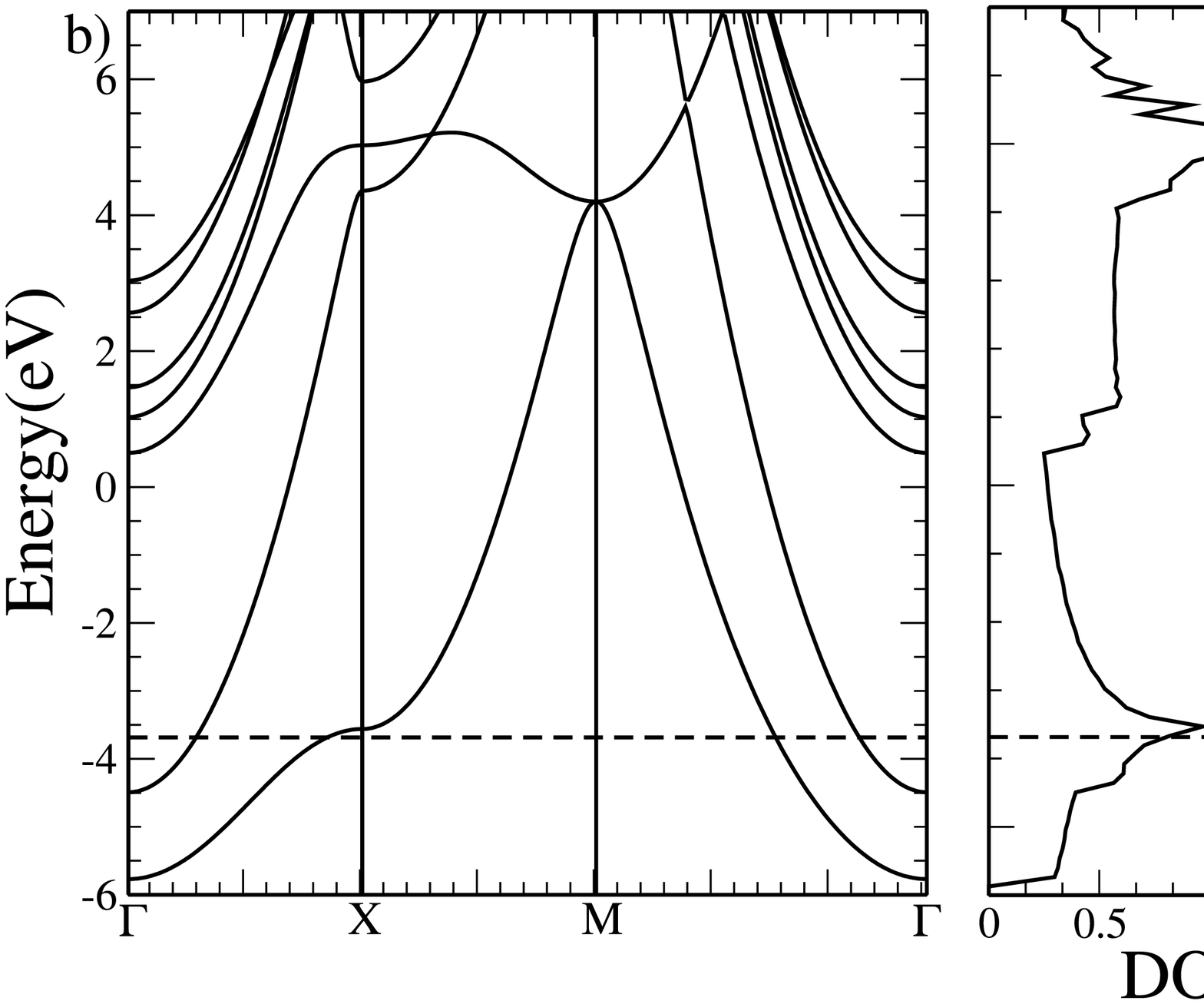}
\caption{Band structure and density of states (DOS) for (a) a hexagonal lithium ML at equilibrium ($r_s=3.02$ a.u.) and 
(b) a square structure with $r_{s}=2.15$ a.u. The Fermi energy is represented by the dotted line.}
\label{fig:Fig2}
\end{figure}

In Fig. \ref{fig:Fig2} we plot the band structure and density of states (DOS) of a lithium ML corresponding
to the hexagonal structure at equilibrium and the square structure with $r_{s}= 2.15$ a.u., just after 
the structural transition. At equilibrium (Fig. 2a) only one band is occupied, which is characterized by
a parabolic-type dispersion of $s$ character, as corresponds to a good NFE-like system.  
For occupied energies the DOS is almost constant, which reflects the quasi two-dimensional 
character of the ML. However, at higher pressures (Fig. 2b) the first band flattens associated to 
the increasing of the band gap at the zone boundary, indicating a clear electronic localization. The energy 
difference between the first two bands at $\Gamma$ decreases so much that the $p_{z}$ band, antisymmetric in the direction perpendicular to the ML, starts to be occupied and opens a new conduction channel. It is also interesting to note that the van Hove singularities lay close to the Fermi energy which, therefore, rises the DOS. All these interesting characteristics can only be explained in terms of the non-local 
character of the ionic pseudopotential \cite{aitor}.\\

\begin{figure}[h!]
\includegraphics[scale=0.325]{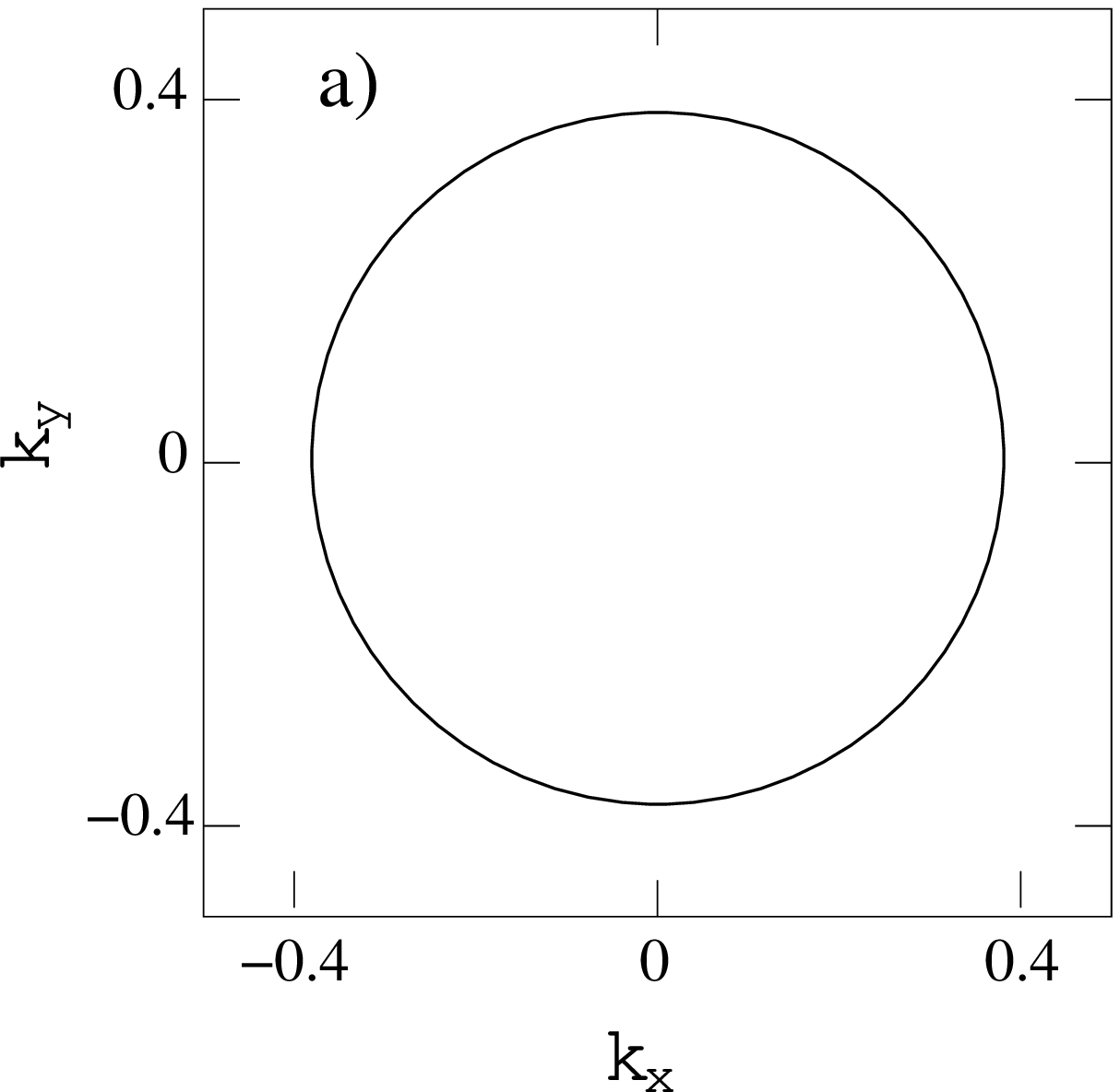}\hspace{0.1cm}
\includegraphics[scale=0.325]{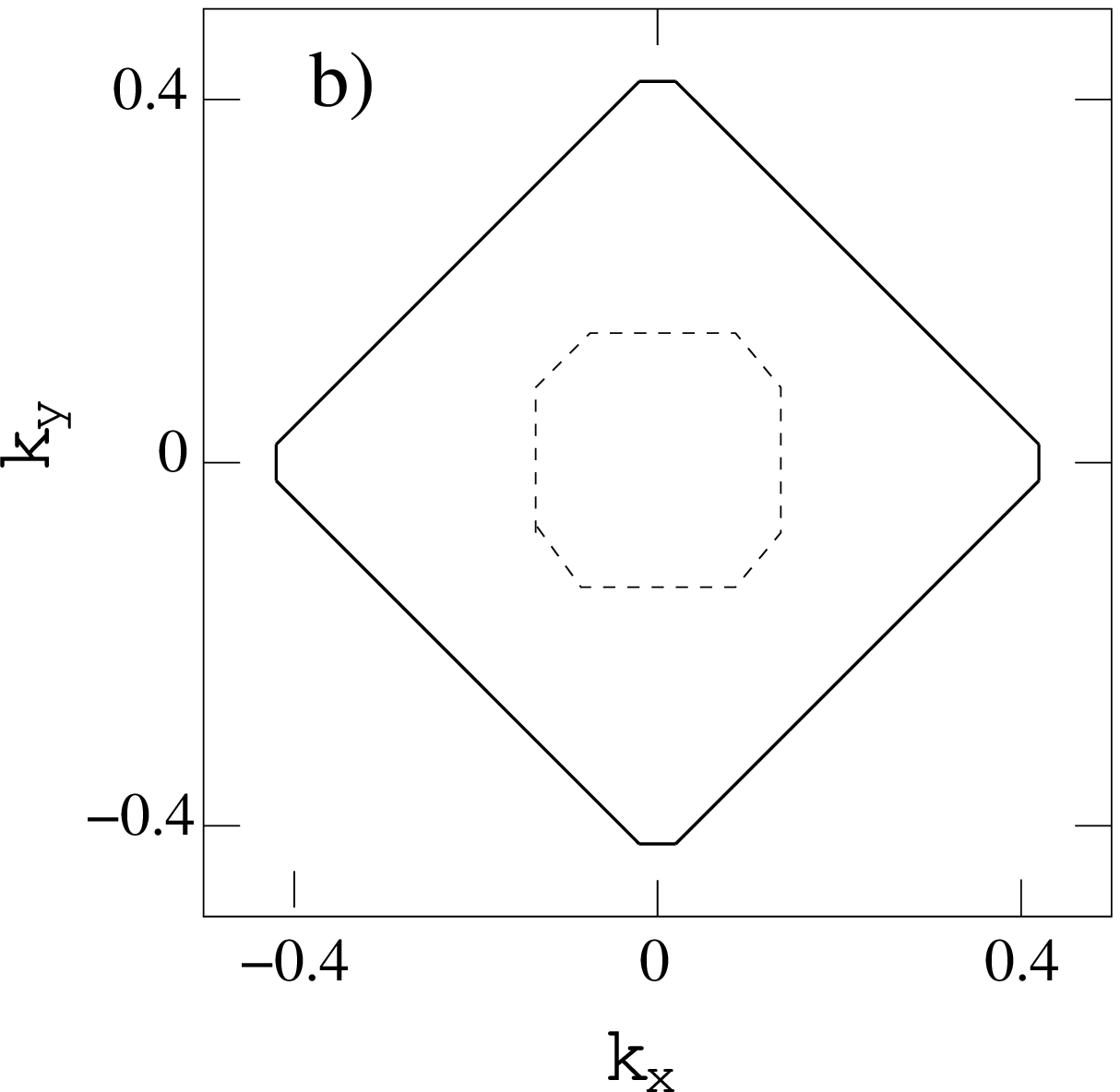}
\caption{Fermi {\it lines} of a) a hexagonal lithium ML at equilibrium and b) a square ML with $r_{s}=2.15$ a.u. The solid line corresponds to the Fermi {\it line} of the $s$ band and the dotted line is associated to the $p_z$ band, which becomes occupied at high densities.}
\label{fig:Fig3} 
\end{figure}

Another essential feature determining electronic properties of the ML is the geometry associated to the 
Fermi {\it line}, which is displayed in Fig. \ref{fig:Fig3} for the same densities and structures as above. At equilibrium just one Fermi  {\it line} exists, associated to the $s$ band, which is almost circular as corresponds to a NFE approximation. However, as mentioned above, this simple picture  breaks at higher densities. At $r_{s}=2.15$ a.u. there are two Fermi {\it lines} 
associated to the two occupied $s$ and $p_z$ bands and, surprisingly, the Fermi {\it line} associated to the $s$ band becomes a perfect square as corresponds to a half filled Hubbard-type model. As it is well known, this model becomes appropriate to describe electronic properties of systems with localized electronic states, just being the opposite case to the NFE approximation valid at normal pressures. The perfect nesting observed at $r_{s}=2.15$ a.u. strongly couples electronic states close to the Fermi {\it line} along the $\overline{\Gamma M}$ direction, and might induce structural Peierls transitions.\\

\subsection{Peierls Transition}

As a consequence of the nesting in the Fermi \textit{line} at $r_{s}\simeq 2.15$ a.u., any 
perturbation which breaks the electronic degeneration might be favored. 
Although different symmetry breaking perturbations might 
be proposed, in this section we have included the analysis of the two simplest diatomic square structures: $dsq_{1}$ and 
$dsq_{2}$. In both structures we choose a diatomic square unit cell and allow the central atom to move along two highly 
symmetric directions: the nearest neighbor ($dsq_{1}$) and second nearest neighbor ($dsq_{2}$) directions.\\

\begin{figure}[h!]
\includegraphics[width=\linewidth,clip=true]{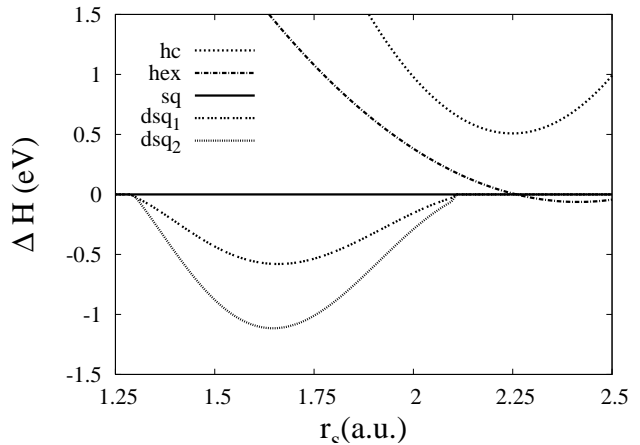}
\caption{Entalphy as a function of the density parameter, $r_{s}$, for all the structures considered: $hc$, $hex$, $sq$, $dsq_1$ and $dsq_2$. All the enthalpies are referenced to the square enthalpy, $\Delta H=H-H_{sq}$. Although the $sq$ structure becomes favored at $r_s=2.25$ a.u., the proposed new diatomic
 square structure ($dsq_2$), where the atom at the center is allowed to move along the direction of the second nearest neighbor, is preferred at higher densities (1.3 a.u.$< r_s <$2.15 a.u.). } 
\label{fig:Fig4}
\end{figure}

In Fig. \ref{fig:Fig4} we present the enthalpies as a function of the density 
for the five structures considered: $hc$, $hex$, $sq$, $dsq_1$ and $dsq_2$. 
It is interesting to note that at $r_{s} \simeq 2.15$ a.u, just the density where the nesting in the Fermi \textit{line} 
of the monoatomic square structure ($sq$) is found, the new $dsq_{2}$ structure proposed in 
this work is favored over the others. Therefore, the perturbation caused by the nesting in the 
Fermi {\it line} consists in a relative movement of the atoms in the second nearest neighbor direction, producing a charge 
density wave in this direction. However, at much higher densities ($r_s=1.35$ a.u.) the $sq$ structure is 
favored again over the diatomic structures.\\

\begin{figure}[h]
\vspace*{0.5cm}
\includegraphics[height=\linewidth,bb=85 33 573 714,angle=-90]{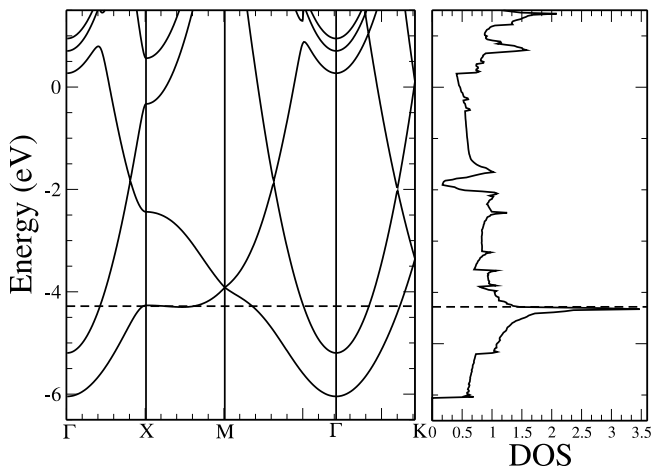}
\caption{Band structure and DOS of a lithium ML with $dsq_2$ structure at $r_s=2.0$ a.u. The $s$ band becomes extraordinary flat in the $\overline{XM}$ direction.}
\label{fig:Fig5}
\end{figure}

\begin{figure}[h]
\centering
\includegraphics[height=0.85\linewidth]{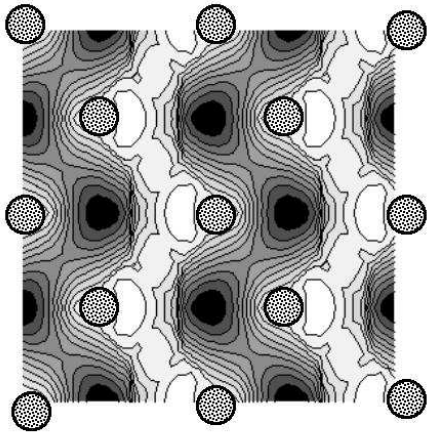}
\caption{LDA valence charge density of the proposed $dsq_2$ lithium ML at $r_s=2.0$ a.u. The ions with core electrons are represented by circles with dots inside, while the valence charge density maximum (black areas) accumulates between them forming a clear one-dimensional conducting channel.}
\label{fig:Fig6}
\end{figure}

Fig. \ref{fig:Fig5} shows the band structure and DOS for the $dsq_2$ structure 
at $r_s=2.0$ a.u., which confirms the clear deviation from a NFE behavior, already manifested for the monoatomic 
structures of a lithium ML. The $s$ band amazingly flattens in the $\overline{XM}$ direction, which indicates a strong electronic 
localization in this direction, although the ML remains metallic. In Fig. \ref{fig:Fig6} we plot the LDA valence charge 
density of the $dsq_2$ structure at $r_s=2.0$ a.u. The charge density accumulates between the ions forming a 
clear one-dimensional conducting channel, which also originates a very sharp singularity in the electronic DOS close 
to the Fermi energy (Fig. \ref{fig:Fig5}).\\

\begin{figure}[h!]
\vspace*{0.5cm}
\includegraphics[width=\linewidth,clip=true]{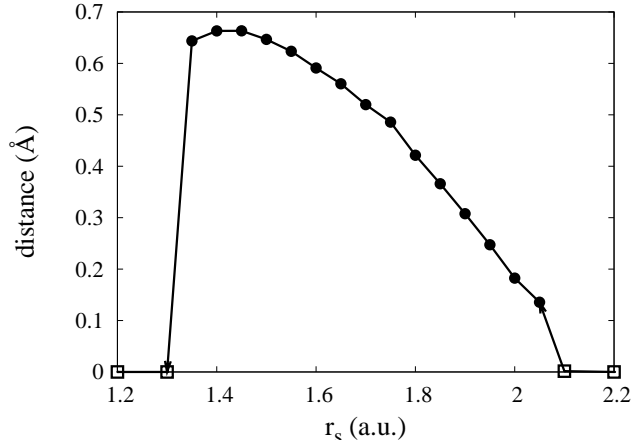}
\caption{Displacement of the central atom in the $dsq_2$ structure  (solid dots) along the second nearest neighbor direction as a function of the density. The open squares correspond to the $sq$ structure, where the atom remains at the center.}
\label{fig:Fig7}
\end{figure}

In order to characterize the phase transition involving the $dsq_2$ structure we have analyzed the displacement of the 
central atom along the minimum energy direction as a function of the density. As it is shown in Fig. 
\ref{fig:Fig7} and  expected from our analysis above, the atomic displacement is zero for $r_{s}>2.15$ a.u. 
However, at $r_{s}=2.15$ a.u. the high anharmonic potential of the central atom induces the presence of a soft phonon 
mode which moves the atom at the center. As it is displayed in Fig. \ref{fig:Fig7}, around $r_s=2.15$ a.u. the 
displacement of the central atom grows continuously with increasing density which indicates the second-order character 
of the $sq \to dsq_2$ transition. However, if we keep rising the density, we find that at $r_s \simeq 1.3$ a.u. the 
displacement of the central atom vanishes discontinuously and the ML goes back to the monoatomic $sq$ structure. We have 
seen that this transition is explained by the presence of an absolute minimum energy at the center while relative minima 
at finite displacements from the center still remain, characterizing the $dsq_2 \to sq$ transition as 
first-order.\\

\subsection{Magnetic Instability}

\begin{figure}[h!]
\includegraphics[width=\linewidth,clip=true]{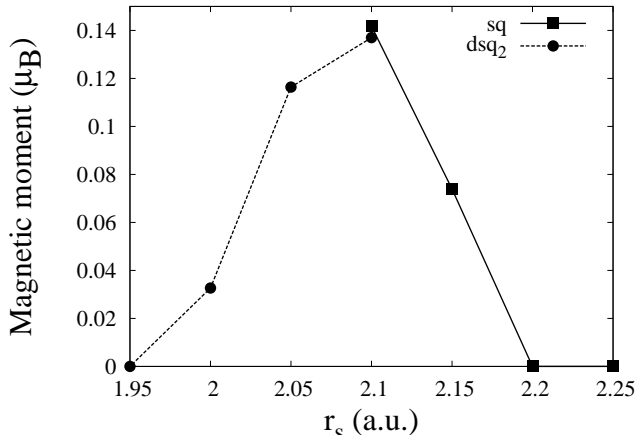}
\caption{Magnetic moment per atom as a function of the density, $r_s$, for a lithium ML  with $sq$ (squares) and $dsq_2$ (dots) structures, which minimize the energy at the pressure range considered. Lithium ML presents a stable ferromagnetic ground state in the proximities of the Peierls transition, with a maximum net magnetic moment per atom of $\sim 0.14 \mu_{\rm B}$, located close to the density where the structural transition ($sq \to dsq_2$) has been predicted, $r_{s}=2.1$ a.u. }
\label{fig:Fig8}
\end{figure}

In a low density electron gas it is well known that a ferromagnetic instability can be induced 
as the kinetic energy penalty in the ferromagnetic state is compensated by the gain associated to the electronic 
exchange \cite{Bloch}. So that, in principle, we would not expect such a transition to occur for high densities. However, as we can see in Figs.  \ref{fig:Fig2} and \ref{fig:Fig5}, the second band starts to be 
occupied as density is increased. Therefore, the effective low electronic density in the $p_{z}$ band, in conjunction with the high DOS in the proximities of the Fermi energy and the nesting observed in the Fermi \textit{line}, could induce a ferromagnetic instability in a lithium ML . Fig. \ref{fig:Fig8} displays \textit{ab initio} calculations within the LSDA approximation which predict that a ferromagnetic instability can occur in a narrow density range: 2.2 a.u.$ > r_s >$ 1.95 a.u. It is interesting to 
notice that the maximum value of the net magnetic moment per conduction electron, $\sim 0.14 \mu_{\rm B}$, is observed 
at $r_{s}=2.1$ a.u, just at the density where the $sq \to dsq_2$ Peierls structural transition takes place. This magnetic instability in a lithium ML, which has been shown to be correlated with electronic and 
structural transitions presented above, also proves the $complexity$ that pressure induces in these, otherwise considered, $simple$ systems.

\section{Conclusions}
\label{sec:C}

In summary, according to the {\it ab initio} calculations presented in this work, although the NFE model correctly describes 
the properties of the lithium ML at normal conditions, it loses its validity as the electronic density increases. We have 
analyzed pressure induced $complexity$ in structural, electronic and even magnetic properties of the ML, finding an 
interesting correlation between them. As density goes up, a flattening of the $s$ band is observed, indicating an 
electronic localization. The  corresponding nesting in the Fermi {\it line} and van Hove singularities in the 
proximities of the Fermi energy observed at $r_{s}\simeq 2.1$ a.u. induce a Peierls second-order structural 
transition ($sq \to dsq_2$). Beyond these features, at around the same density, the non-local character of the 
ionic pseudopotential allows to start filling the $p_z$ band, and its low effective initial occupation causes a 
ferromagnetic Bloch instability. Besides of the intrinsic interest of studying the physical properties of systems with 
reduced dimensionality, the simplicity of the ML allows us to make a more detailed analysis and by extending these conclusions to bulk systems will provide us a very useful  perspective to understand the physical origin of the {\it complexity} observed in compressed bulk lithium \cite{alv1}.\\

\begin{acknowledgments}
A. Rodriguez-Prieto would like to acknowledge financial support from the Donostia International Physics Center (DIPC) and the Basque Hezkuntza, Unibertsitate eta Ikerketa Saila.
\end{acknowledgments}

\bibliography{article}

\begin{thebibliography}{26}
\bibitem[{rs()}]{rs}
\bibinfo{note}{$r_{s}$ is the two-dimensional linear density parameter, defined
  by the relation $A/N=\pi(r_{s}a_{0})^{2}$ where $N$ is the number of nuclei
  in a ML with area $A$ and $a_{0}$ is the Bohr radius.}
\end{thebibliography}

\end{document}